\documentclass[%
reprint,
showpacs,
 amsmath,amssymb,
 aps,
]{revtex4-1}

\usepackage{graphicx}
\usepackage{dcolumn}
\usepackage{bm}
\usepackage{color}
\usepackage{braket}

\begin{document}

\title{Parametric Resonances and Resonant Delocalization in Quasi-Phase Matched Photon-pair Generation and Quantum Frequency Conversion}

\author{Philip B. Main}
  \email{P.B.Main@bath.ac.uk}
 \affiliation{
  Centre for Photonics and Photonic Materials, 
  Department of Physics, University of Bath, Bath BA27AY, UK}
  
 \author{Peter J. Mosley}
 \affiliation{
 Centre for Photonics and Photonic Materials, 
 Department of Physics, University of Bath, Bath BA27AY, UK}  

\author{Andrey V. Gorbach}
  \email{A.Gorbach@bath.ac.uk}
  \affiliation{
  Centre for Photonics and Photonic Materials, 
  Department of Physics, University of Bath, Bath BA27AY, UK}

\date{\today}

\begin{abstract}
The existing widely-accepted theory of photon-pair generation via spontaneous down-conversion (SPDC) in nonlinear optical crystals and waveguides is incomplete, as it fails to account for the important physical phenomenon of parametric resonances. We demonstrate that exponential gain of classical fields in the regime of parametric resonance corresponds to resonant delocalization in the Glauber-Fock model of quantum SPDC. We propose a quantitative measure of localisation of Floquet eigen-modes as an analogue of classical gain to identify regimes of resonant delocalization. Using this method, we are able to reconstruct the classical "Arnold tongues" map of domains of instabilities for SPDC. We also predict novel regimes of resonant delocalization  in the two-level model describing quantum frequency conversion processes.
\end{abstract}

\maketitle



\section{\label{sec:Intro}Introduction}

Parametric resonances are a well-known instability mechanism triggered by periodical modulation of a system parameter. A classical example is an oscillator with periodically modulated eigenfrequency, whose dynamics are governed by the renowned Mathieu equation \cite{Arnold1989}. The important distinct feature of parametric resonances is the existence of multiple frequency ranges of instability, even when the modulation is purely harmonic. Furthermore, positions and bandwiths of the instability regions change with the modulation amplitude. Parametric resonances govern a wide range of physical phenomena, including pattern formation in liquids on a vibrating substrate \cite{Faraday1831}, periodically forced reaction-diffusion systems \cite{Lin2000}, Bose-Einstein condensates with modulated interactions \cite{Staliunas2002, Engels2007}, multi-mode lasers \cite{Szwaj1998}, and unstable vibrations of London's Millennium bridge \cite{Piccardo2008}. In nonlinear optical systems, similar parametric instabilities arise from spatial modulation of dispersive \cite{Conforti2014}, dissipative \cite{Tarasov2016, Perego2018} or nonlinear \cite{Abdullaev1997, Staliunas2013} properties of the medium.

In a different context, modulation of nonlinearity along the path of interacting optical waves is a well-known technique for effective compensation of their momentum mismatch knows as quasi-phase matching (QPM) \cite{Armstrong1962}. In particular, periodic alternation of the sign of $\chi_2$ nonlinearity has become a widely recognised technique for efficient second harmonic generation in bulk crystals and waveguides \cite{Feng1980, Jundt1991, Mizuuchi1992, Bromberg1992}. Later QPM has also been adapted for optical parametric oscillation, parametric amplification \cite{Bortz1995, Myers1995} and spontaneous parametric down conversion (SPDC) processes \cite{Tanzilli2001b,Banaszek2001,Xu2012d}. The latter form the basis of one of the most promising and robust schemes of generation of correlated photon pairs for applications in quantum computing, metrology, and development of heralded single photon sources \cite{Burnham1970b,Eisaman2011,Couteau2018}. Despite apparent similarities, the relationship between parametric instabilities and QPM-driven parametric processes has not been fully explored. 

\begin{figure}
    \centering
    \includegraphics[width=0.45\textwidth]{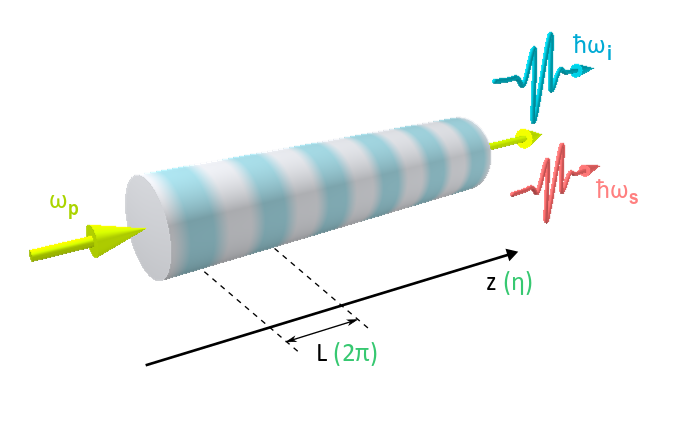}
    \caption{Scheme of SPDC process in a $\chi_2$ waveguide with modulated nonlinearity.}
    \label{fig:SPDC_scheme}
\end{figure}

In this work we demonstrate that the conventional quantum-mechanical treatment of photon-pair generation by SPDC fails to capture parametric resonances, in contrast with the classical model. We reveal the intrinsic connection between parametric resonances and the phenomenon of resonant delocalization in Glaube-Fock lattices, and obtain the quantum SPDC analogue of the classical "Arnold tongues" picture of resonance domains. We furthermore explore this connection to predict novel regimes of resonant delocalization and Rabi oscillations in the two-level model, describing sum- and difference-frequency generation. For clarity, we focus our discussion on the case of nonlinear interactions in one dimension in a material with modulated $\chi_2$ nonlinearity, as shown in Fig.~\ref{fig:SPDC_scheme}. The archetypal example of this is a single-mode waveguide periodically poled to achieve QPM, however our analysis can be straightforwardly extended to any $\chi_2$ or $\chi_3$ material exhibiting periodic modulation of its nonlinearity.

\section{\label{sec:SPDC_and_psi2}Photon pair generation and two-photon state function in SPDC}

In a $\chi_2$-driven SPDC process, a higher energy photon (pump, $\omega_p$) from a bright source is spontaneously converted into a pair of lower energy photons (signal and idler, $\omega_s+\omega_i=\omega_p$), see Fig.~\ref{fig:SPDC_scheme}. In a waveguide, all photons propagate in the same direction and the important parameter which ultimately determines the properties of the generated signal-idler pairs, for example their joint frequency structure, is the momentum (propagation constant) mismatch of the interacting waveguide modes \cite{Main2016}: 
\begin{equation}
\Delta\beta(\omega_s,\omega_i)=\beta_p(\omega_s+\omega_i)-\beta_s(\omega_s)-\beta_i(\omega_i)\;. 
\label{eq:dBeta}
\end{equation}
While it is often not possible to achieve direct phase matching $\Delta\beta=0$ for a desired combination of frequencies and waveguide modes, the QPM technique relies on modulation of the interaction strength along the waveguide to effectively compensate a non-zero mismatch. Treating the pump field classically, and neglecting pump depletion, the spontaneous creation of signal-idler photon pairs can be described by the following interaction Hamiltonian \cite{Mandel1995, Xu2012d}:
\begin{equation}
\hat{H}_I = \tilde{\gamma}(\eta) [ e^{-iR \eta} \hat{a}^\dagger_s \hat{a}^\dagger_i + e^{iR \eta} \hat{a}_s \hat{a}_i]\;,
\label{eq:Hint}
\end{equation}
where $\eta=\kappa z$ is the dimensionless propagation distance related to the modulation period $L=2\pi/\kappa$, $\tilde{\gamma}(\eta)=\sqrt{P_0}\gamma(\eta)/\kappa$ is the effective interaction which encapsulates the modulated waveguide nonlinearity $\gamma(\eta+2\pi)=\gamma(\eta)$ and the pump power $P_0$, $R=\Delta\beta/\kappa$ is the ratio between the momentum mismatch and the reciprocal modulation period. Hence, setting vacuum state $\ket{\textrm{vac}}$ as the initial condition at $\eta=0$, the state vector is given by:
\begin{equation}
\ket{\psi}(\eta)=\textrm{exp}\left[-i\int_0^\eta \hat{H}_I(\eta^\prime)d\eta^\prime \right] \ket{\textrm{vac}} \;.
\label{eq:psi_gen}
\end{equation}

The next commonly-used step is to apply a perturbation expansion of the exponential term in the above expression, assuming a weak interaction \cite{Rubin1996,Grice1997a,Yang2008a}:
\begin{equation}
\ket{\psi}(\eta)\approx\left[1-i\int_0^\eta \hat{H}_I(\eta^\prime)d\eta^\prime +\dots \right] \ket{\textrm{vac}} \;,
\label{eq:psi_expansion}
\end{equation}
which naturally leads to the decomposition of the state into single- and multiple photon-pair terms: $\ket{\psi}(\eta)\approx\ket{\textrm{vac}} +\ket{\psi_2}+\ket{\psi_4}+\dots$. In particular, from Eq.(\ref{eq:psi_expansion}) the two-photon state is obtained:
\begin{eqnarray}
\ket{\psi_2}(\eta) =  \left(-i\int_0^\eta \tilde{\gamma}(\eta^\prime) e^{-iR \eta^\prime}d\eta^\prime \right) \hat{a}^\dagger_s \hat{a}^\dagger_i \ket{\mathrm{vac}} 
\label{eq:psi2}
\end{eqnarray}

Expanding the $2\pi$-periodic interaction function in the Fourier series:
\begin{equation}
\tilde{\gamma}(\eta)=\sum_m \tilde{\gamma}_j e^{im\eta}\;,
\end{equation}
it is easy to see that the two-photon function amplitude grows linearly with propagation distance if $R$ is integer. In other words, photon pair generation occurs when the momentum mismatch $\Delta\beta$ coincides with the reciprocal period of $m$-th harmonic $m\kappa$ of the nonlinearity modulation function $\gamma(\eta)$. 

We emphasise, that for the case of simple harmonic modulation: 
\begin{equation}
\tilde{\gamma}(\eta)=\tilde{\gamma}_0 \cos(\eta)\;,
\label{eq:gam_harmonic}
\end{equation}
according to the well-known in literature result for the two-photon function in Eq.~(\ref{eq:psi2}), the growth of the two-photon state amplitude is only observed when $R=\pm 1$, i.e. when $\Delta\beta=\pm\kappa$. Furthermore, this result does not depend on the amplitude of the modulation $\widetilde{\gamma}_0$.

\section{Parametric resonances in classical parametric amplification}

\begin{figure}
    \centering
    \includegraphics[width=0.4\textwidth]{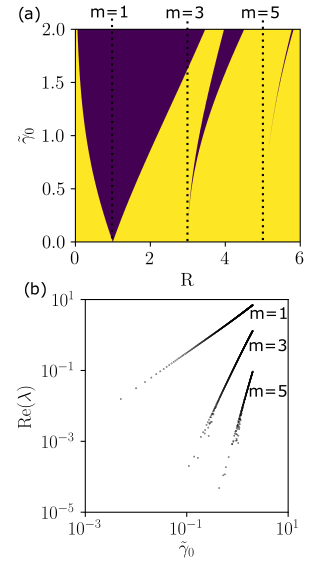}
    \caption{Parametric resonances in the classical down-converted fields: a) map of unstable regions ("Arnold tongues") of Eq.~(\ref{eq:classical_single_variable}), indicated with dark (blue) colour, on the plane of parameters $(R,\tilde{\gamma}_0)$; b) Maximal gain per period within $m=1,3,5$ "tongues" as function of the interaction strength.}
    \label{fig:classical_gain}
\end{figure}

Let us now consider the classical analogue of the SPDC process, i.e. the process of parametric amplification. Under the same assumption of undepleted pump as used in derivation of the Hamiltonian in Eq.~(\ref{eq:Hint}), the interacting (weak) signal and idler field amplitudes $A_{s,i}$ evolve along the waveguide length according to \cite{agrawal}:
\begin{eqnarray}
\frac{d A_s}{d \eta} = i\tilde{\gamma}(\eta) e^{-iR\eta} A_i^* \;, \qquad \frac{d A_i}{d \eta} = i\tilde{\gamma}(\eta) e^{-iR\eta} A_s^* \;.
\label{eq:classical}
\end{eqnarray}
Making the substitution $X =  [A_s + A_i] \exp\left(i R \eta/2\right)$, and using the simple harmonic modulation in Eq.~(\ref{eq:gam_harmonic}),  the above system can be casted into a Mathieu-type oscillator equation:
\begin{equation}
i \frac{d X}{d \eta} + \frac{R}{2} X + \tilde{\gamma}_0 \cos [\eta] X^* = 0 \;.
\label{eq:classical_single_variable}
\end{equation}
To analyse dynamics of this ODE with periodically varying coefficients, it is convenient to consider the corresponding Floquet operator, which maps the field over one period: $[X(\eta+2\pi),X^*(\eta+2\pi)]^T=\hat{F}\cdot[X(\eta),X^*(\eta)]^T$. The operator $\hat{F}$ can be constructed numerically by integrating Eq.~(\ref{eq:classical_single_variable}) with two orthogonal initial conditions. Spectral properties of $\hat{F}$ determine stability of the system in Eq.~(\ref{eq:classical_single_variable}):
\begin{equation}
\hat{F}\cdot \vec{\nu}^{(n)}=\lambda_n \vec{\nu}^{(n)}\;. 
\label{eq:F_eigen_problem}
\end{equation}
An eigenvalue $\lambda_n$ with a positive real part corresponds to exponential gain in signal/idler fields. In Fig.~\ref{fig:classical_gain}(a) the corresponding gain regions are indicated on the plane of parameters $(R,\tilde{\gamma_0})$, and have the typical "Arnold tongues" structure known for solutions to the Mathieu equation and seen in other systems exhibiting parametric resonance \cite{Arnold1989}. For a fixed interaction strength $\tilde{\gamma}_0$ the system is unstable within multiple regions of $R$. These regions emerge from the set of points $R=m$, $m=\pm1,\pm3,\pm5,\dots$ on $\tilde{\gamma}_0=0$ axis, expanding and shifting as $\tilde{\gamma}_0$ increases. In Fig.~\ref{fig:classical_gain}(b) the maximal gain, i.e. real part of eigenvalues $\lambda$, as function of interaction strength is plotted for the first three "tongues" ($m=1,3,5$). It scales as $\tilde{\gamma}_0^m$, consistent with Arnold's scaling law \cite{Arnold1983, Ecke1989}.

\section{Parametric resonances and resonant delocalization in SPDC}

The analysis above reveals a fundamental inconsistency between classical theory and the approximation commonly used in the quantum-mechanical treatment of QPM down-conversion shown in Eq.~(\ref{eq:psi_expansion}). It is easy to see that with the interaction Hamiltonian defined in Eq. (\ref{eq:Hint}), equations for $\hat{a}_s$ and $\hat{a}_i$ operators in the Heisenberg picture have similar structure to Eqs.~(\ref{eq:classical}). Therefore one should expect to observe growth of signal and idler photon pair numbers in the parameter regions where classical model predicts parametric amplification. However, neither the existence of higher order resonances ($m=\pm3,\pm5, \dots$), nor the resonance bandwidth and position dependencies on the modulation strength are reflected in the two-photon function amplitude in Eq.~(\ref{eq:psi2}) with the simple harmonic modulation in Eq.~(\ref{eq:gam_harmonic}). The inclusion of higher-order expansion terms in Eq.~(\ref{eq:psi_expansion}) does not restore any of these well-known parametric resonance features. Apparently, the widely adapted perturbation expansion procedure in Eq.~(\ref{eq:psi_expansion}) fails to capture the important physical aspects of SPDC processes, and needs to be reconsidered. 

To develop an analogue of the classical Floquet analysis for the SPDC process, we adapt the Fock basis of signal-idler photon pairs $\{ \ket{\psi_{n}} = \ket{nn} e^{-i n R \eta} \}$, where $\ket{nn}=(a_s^\dagger a_i^\dagger)^n \ket{vac}$.
This allows the signal and idler creation and annihilation operators to be absorbed into outer products. Hence, the interaction Hamiltonian in Eq.~(\ref{eq:Hint}) becomes:
\begin{equation}
H_I = \tilde{\gamma}(\eta) \sum_n\left[(n + 1)  \ket{\psi_{n+1}}\bra{\psi_{n}} + n \ket{\psi_{n-1}}\bra{\psi_{n}}\right]
\label{eq:Hint_fock}
\end{equation}
and the evolution of the state vector $\ket{\psi} = \sum U_n \ket{\psi_n}$ is governed by the set of ODEs with periodic coefficients:       
\begin{equation}
-i\frac{d U_n}{d \eta} = \tilde{\gamma}(\eta) \left[ n U_{n-1} + (n+1)U_{n+1} \right] - n R U_n
\label{eq:lattice}
\end{equation}
The corresponding Floquet operator can be obtained by taking the product of a semi-infinite set of linearly independent solutions of the above system integrated over one modulation period: $\hat{F} = \phi_i \otimes \phi_i$. This was done numerically with the help of ODEPACK automated Adams/BDF ODE integrator \cite{Hindmarsh1982}. Evolution of an arbitrary initial state is then obtained by repeated translations with $\hat{F}$. In numerical modelling the semi-infinite system in Eq.~(\ref{eq:lattice}) was manually truncated at large enough $n$, such that no boundary effects are observed in propagation of initial vacuum state over $10^4$ periods.

\begin{figure}
\includegraphics[width = 0.48\textwidth]{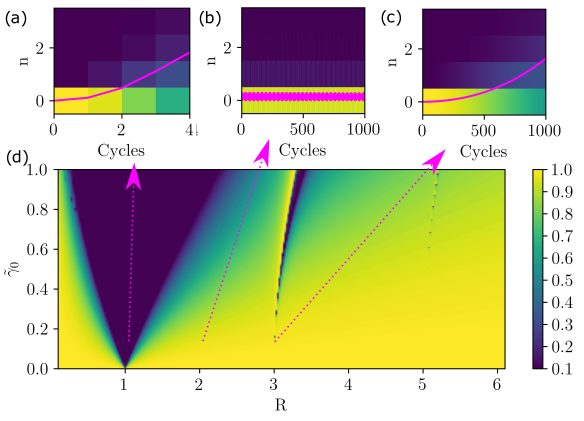}
\caption{Parametric resonances in quantum SPDC: (a),(b),(c) Evolution of the state vector in terms of Fock states amplitudes $|U_n|^2$ for ratios $R  = 1, 2, 3$ respectively with $\ket{\psi}(0)=\ket{vac}$ and $\tilde{\gamma}_0=0.2$. Pink lines show the corresponding average number of photons $\bra{\psi}\hat{n}_s\ket{\psi}$; d) "Arnold tongues" generated from the localization parameter $\mathcal{P}(R, \tilde{\gamma})$ of the Floquet eigen-modes. Resonant coupling between states occurs in the dark blue regions where $\mathcal{P}$ is small. The colourbar applies to all four plots. }
\label{fig:spdc_tongues}
\end{figure}

Unlike its classical counter-part in Eq.~(\ref{eq:classical}), the system in Eq.~(\ref{eq:lattice}) preserves the norm $\sum_n |U_n|^2$, and therefore cannot have exponentially growing solutions. In Fig.~\ref{fig:spdc_tongues}(a)-(c) the evolution of the state vector is illustrated for the case of simple harmonic modulation $\tilde\gamma(\eta)$ in Eq.~(\ref{eq:gam_harmonic}), with $R=1,3,5$, respectively,  and initial vacuum state $\ket{\psi}(0)=\ket{\textrm{vac}}$. Two qualitatively different types of evolution are observed for $R=1,3$ and $R=2$ cases. In $R=2$ case (no parametric resonance in classical system), Fig.~\ref{fig:spdc_tongues}(b), a partial beating between the vacuum and higher order terms is observed. In contrast, in $R=1,3$ cases (parametric resonances), Fig.~~\ref{fig:spdc_tongues}(a) and (c), the system gradually evolves into the pairwise-correlated thermal state. The total number of signal and idler photons $\bra{\psi}\hat{n}_s\ket{\psi}$  grows in this process, which corresponds to the exponential explosion of the classical field intensities. The characteristic lengthscales of resonant coupling dynamics in $R=1$ and $R=3$ cases are different by three orders of magnitude, which is in agreement with the scaling law of parametric resonances, cf. Fig.~\ref{fig:classical_gain}(b).

The system in Eq.~(\ref{eq:lattice}), also known as the Glaube-Fock lattice \cite{Perez-Leija2012}, is equivalent to a semi-infinite 1D Bloch lattice of coupled detuned oscillators, $n$th oscillator having eigen-frequency $nR$, with inhomogeneous and periodically varying coupling. The modulation of coupling enables effective cross-talk between the detuned oscillators, leading to the so-called resonant delocalization \cite{Perez-Leija2012}. The phenomenon is known for the $R=\pm1$ case \cite{Perez-Leija2012, Sukhorukov2013a}, however the present theory in Ref. \cite{Perez-Leija2012} fails to predict higher order parametric resonances $R=\pm3,\pm5,\dots$. 

While gain is replaced by resonant delocalization, the spectrum $\lambda_n$ of $\hat{F}$ no longer carries any information about such resonances. Instead, the structure of its eigen-modes $\vec{\nu}^{(n)}$ needs to be analysed. For this purpose, we introduce a measure of localization of Floquet eigen-modes, similar to the so-called inverse participation ratio used e.g. for studies of Anderson localization in lattices \cite{Izrailev2012}. While in the SPDC process coupling to the vacuum state plays crucial role, we define the localization parameter as $\mathcal{P} = \sum_n (\nu_{0}^{(n)})^4$, where $\nu_0^{(n)}$ is the first (vacuum) component of the $n$-th normalized eigen-mode. In the limit of weak interaction, eigen-modes of $\hat{F}$ converge to Fock states, i.e. $n$th eigen-mode is localized on the respective lattice site. It is easy to see that in this limit the localization parameter tends to its maximal value $\mathcal{P}\to 1$. In the opposite limit of strong interaction, we expect all eigen-modes to be equally spread across the lattice, so that $(\nu_0^{(n)})^2\sim 1/N\; \forall \; n$, where $N$ is the size of the truncated Glaube-Fock lattice. In this limit the localization parameter tends to its minimal value $\mathcal{P}\to 1/N$. In Fig.~\ref{fig:spdc_tongues}(d) we plot $\mathcal{P}$ for the system in Eq.~(\ref{eq:lattice}) as function of the modulation parameter $R$ and interaction strength $\tilde{\gamma}_0$. We observe several distinct regions of low $\mathcal{P}$, which form the well-known classical picture of "Arnold tongues", cf. Fig.~\ref{fig:classical_gain}(a), and correspond to the resonant delocalization regime. Remarkably, our analysis predicts higher order resonances, in full correspondence with the classical model.

The perturbation solution in Eqs.~(\ref{eq:psi_expansion}), (\ref{eq:psi2}) is recovered by assuming the hierarchy of smallness of Fock state amplitudes: $|U_0|\gg|U_1|\gg|U_2|\dots$. In this regime the system in Eq.~(\ref{eq:lattice}) becomes:
\begin{equation}
\left\{
\begin{array}{l}
-i(dU_0/d\eta)=0\;,\qquad U_0=1\;,\\
-i(dU_1/d\eta)=-RU_1+\tilde{\gamma}(\eta)U_0\;,\\
-i(dU_2/d\eta)=-2RU_2+2\tilde{\gamma}(\eta)U_1\;,\\
\dots
\end{array}
\right.
\label{eq:pert_oscillators}
\end{equation}
By solving the above system recursively, dynamics of each multi-photon state is governed by a simple driven oscillator-type equation. Here, the solution for $U_{n-1}$ from the previous step serves as an effective external driving force in the equation for $U_n$. In other words, in this perturbation expansion procedure parametric resonances are replaced by standard resonances. It is easy to see that the resonance condition is the same for all $U_n$. In particular, for simple harmonic modulation of $\tilde{\gamma}(\eta)$ the above system has only $R=\pm 1$ resonance. Solving Eqs.~(\ref{eq:pert_oscillators}) for $U_1$, the two-photon function in Eq.~(\ref{eq:psi2}) is restored.

\begin{figure}
\includegraphics[width = 0.48\textwidth]{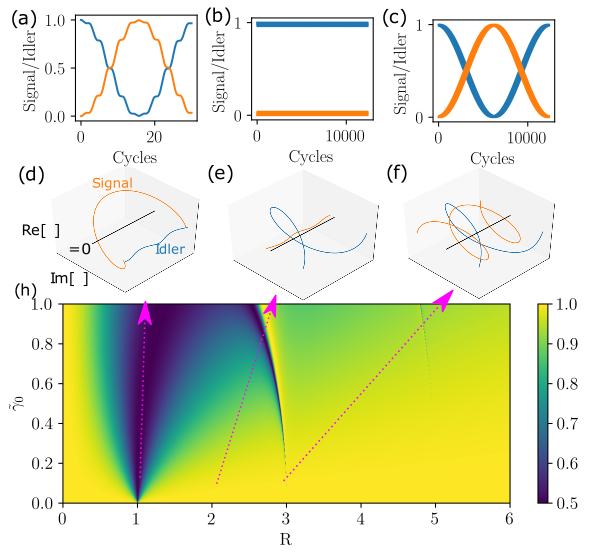}
\caption{Resonant delocalization in quantum frequency conversion: (a)-(c) Dynamics of the system in Eq.~(\ref{eq:classical_DFG}) with initial condition $A_i(0)=1$, $A_s(0)=0$ in the case of simple harmonic modulation in Eq.~(\ref{eq:gam_harmonic}),  $\tilde{\gamma}_0=0.2$,  and $R=1,2,3$, respectively; (d)-(f) structure of the corresponding Floquet eigen-modes. Black line indicates $\nu=0$ axis; (g) "Arnold tongues" of the system generated from the localization parameter $\mathcal{P}$ if the Floquet eigen-modes.}
\label{fig:diff_gen}
\end{figure}

\section{Resonant delocalization in sum- and difference-frequency generation}

We emphasise that resonant delocalization is a generic mechanism which can be observed in a wide range of classical and quantum coupled oscillator-type systems with periodically modulated parameters. It is instructive to consider another type of three-wave mixing process, the so-called difference- and sum-frequency generation, whereby an idler wave (or photon) is injected together with a pump into the waveguide, producing signal at $\omega_s=\omega_i\mp \omega_p$ \cite{agrawal}. In the context of quantum optics such processes are also known as quantum frequency conversion \cite{Kumar1990}. In the undepleted pump approximation, both classical (idler wave) and quantum (idler photon) models of this process are similar to the dynamics of a two level system:
\begin{eqnarray}
\frac{d A_s}{d \eta} =i\tilde{\gamma}(\eta)e^{-iR\eta} A_i \;, \qquad \frac{d A_i}{d \eta} = i\tilde{\gamma}^*(\eta)e^{iR\eta} A_s \;.
\label{eq:classical_DFG}
\end{eqnarray}
In the difference- (sum-) frequency generation case the initial condition is set to $A_i(0)=1$, $A_s(0)=0$ ($A_i(0)=0$, $A_s(0)=1$).
Unlike the model in Eq.~(\ref{eq:classical}), there can be no exponential gain in the above system. Instead, by tuning the model parameters, one can observe a resonant beating between signal and idler, as illustrated in Figs.~\ref{fig:diff_gen}(a)-(c). The observed complete Rabi oscillations in $R=1$ case is well understood. Here, one of the exponents in $\tilde{\gamma}(\eta)\sim\cos(\eta)=0.5(e^{i\eta}+e^{-i\eta})$ modulation cancels the phase-mismatch exponents, thus enabling efficient coupling. However, this simple logic fails to explain similar oscillations in the $R=3$ case. In Fig.~\ref{fig:diff_gen}(d)-(f) the structure of the corresponding Floquet eigen-modes is illustrated (for clarity, only one of the two conjugate modes is shown). In the $R=1$ and $R=3$ cases, both signal and idler components of the eigen-mode retain large amplitudes throughout the modulation period. In contrast, in $R=2$ case one component of the eigen-mode has a much lower amplitude than the other component, therefore signal and idler are practically de-coupled. Adapting the definition of the localization parameter $\mathcal{P}$ for this case through the idler component of eigen-modes, we reveal the "Arnold tongue"-like structure of resonant delocalization regions in the space of parameters $(R,\tilde{\gamma})$, see Fig.~\ref{fig:diff_gen}(g). In full analogy to parametric resonances, the effective strength of the $R=3$ "resonance" is weaker than $R=1$, and the complete frequency conversion is observed over a larger number of modulation cycles, cf. Figs.~\ref{fig:diff_gen}(a) and (c).



\section{Summary}

It is well-known that the couplings between bright optical fields in structures with periodically modulated nonlinearity exhibit parametric resonances. We have shown that photon pair generation via SPDC in a waveguide with single-harmonic modulation of $\chi_2$ nonlinearity can also be observed within multiple domains in the parameter space of modulation strength and period, in agreement with the regimes of instability of the classical Mathieu equations ("Arnold's tongues"). The widely accepted theory of SPDC based on the perturbative derivation of the so-called two-photon function, Eqs. (\ref{eq:psi_expansion}), (\ref{eq:psi2}), fails to predict such resonances. 

We have demonstrated that parametric resonances in SPDC correspond to resonant delocalization in the Glaube-Fock model. Unlike classical parametric amplification, such resonant delocalization is not reflected in the spectrum of the corresponding Floquet operator. Instead, the structure of the Floquet eigenmodes must be analysed. By introducing the corresponding localization parameter $\mathcal{P}$, we have recovered multiple domains of photon-pair generation. However, the localization parameter $\mathcal{P}$ gives no information about the strength of such resonances, unlike the exponential gain parameter calculated for non-Hermitian models.

Our method helps to predict the phenomenon of resonant delocalization in the generic class of coupled oscillator-type models. In particular, we have explored novel regimes of resonant delocalization in the QPM photon frequency-conversion process. The established analogy between parametric resonances and resonant delocalization in parametric down-conversion processes brings a fresh insight into such seemingly unrelated dynamical mechanisms, and can help in developing better tools for their analysis. Furthermore, we note the possibility of harnessing these resonances for previously unexplored phase matching. For example, converting emission from atomic transitions to telecommunications wavelengths typically requires very short poling periods with commensurately tight fabrication tolerances\cite{Wright2018}; exploiting a higher-order parametric resonance would lengthen the poling period required and relax the fabrication requirements.

\bibliography{references}

\begin{thebibliography}{39}%
\makeatletter
\providecommand \@ifxundefined [1]{%
 \@ifx{#1\undefined}
}%
\providecommand \@ifnum [1]{%
 \ifnum #1\expandafter \@firstoftwo
 \else \expandafter \@secondoftwo
 \fi
}%
\providecommand \@ifx [1]{%
 \ifx #1\expandafter \@firstoftwo
 \else \expandafter \@secondoftwo
 \fi
}%
\providecommand \natexlab [1]{#1}%
\providecommand \enquote  [1]{``#1''}%
\providecommand \bibnamefont  [1]{#1}%
\providecommand \bibfnamefont [1]{#1}%
\providecommand \citenamefont [1]{#1}%
\providecommand \href@noop [0]{\@secondoftwo}%
\providecommand \href [0]{\begingroup \@sanitize@url \@href}%
\providecommand \@href[1]{\@@startlink{#1}\@@href}%
\providecommand \@@href[1]{\endgroup#1\@@endlink}%
\providecommand \@sanitize@url [0]{\catcode `\\12\catcode `\$12\catcode
  `\&12\catcode `\#12\catcode `\^12\catcode `\_12\catcode `\%12\relax}%
\providecommand \@@startlink[1]{}%
\providecommand \@@endlink[0]{}%
\providecommand \url  [0]{\begingroup\@sanitize@url \@url }%
\providecommand \@url [1]{\endgroup\@href {#1}{\urlprefix }}%
\providecommand \urlprefix  [0]{URL }%
\providecommand \Eprint [0]{\href }%
\providecommand \doibase [0]{http://dx.doi.org/}%
\providecommand \selectlanguage [0]{\@gobble}%
\providecommand \bibinfo  [0]{\@secondoftwo}%
\providecommand \bibfield  [0]{\@secondoftwo}%
\providecommand \translation [1]{[#1]}%
\providecommand \BibitemOpen [0]{}%
\providecommand \bibitemStop [0]{}%
\providecommand \bibitemNoStop [0]{.\EOS\space}%
\providecommand \EOS [0]{\spacefactor3000\relax}%
\providecommand \BibitemShut  [1]{\csname bibitem#1\endcsname}%
\let\auto@bib@innerbib\@empty
\bibitem [{\citenamefont {Arnold}(1989)}]{Arnold1989}%
  \BibitemOpen
  \bibfield  {author} {\bibinfo {author} {\bibfnamefont {V.}~\bibnamefont
  {Arnold}},\ }\href {\doibase 10.1007/978-1-4757-2063-1} {\emph {\bibinfo
  {title} {{Mathematical methods of classical mechanics}}}}\ (\bibinfo
  {publisher} {New York : Springer-Verlag},\ \bibinfo {year}
  {1989})\BibitemShut {NoStop}%
\bibitem [{\citenamefont {Faraday}(1831)}]{Faraday1831}%
  \BibitemOpen
  \bibfield  {author} {\bibinfo {author} {\bibfnamefont {M.}~\bibnamefont
  {Faraday}},\ }\href@noop {} {\bibfield  {journal} {\bibinfo  {journal}
  {Philosophical Transactions of the Royal Society of London}\ }\textbf
  {\bibinfo {volume} {121}},\ \bibinfo {pages} {299} (\bibinfo {year}
  {1831})}\BibitemShut {NoStop}%
\bibitem [{\citenamefont {Lin}\ \emph {et~al.}(2000)\citenamefont {Lin},
  \citenamefont {Bertram}, \citenamefont {Martinez}, \citenamefont {Swinney},
  \citenamefont {Ardelea},\ and\ \citenamefont {Carey}}]{Lin2000}%
  \BibitemOpen
  \bibfield  {author} {\bibinfo {author} {\bibfnamefont {A.~L.}\ \bibnamefont
  {Lin}}, \bibinfo {author} {\bibfnamefont {M.}~\bibnamefont {Bertram}},
  \bibinfo {author} {\bibfnamefont {K.}~\bibnamefont {Martinez}}, \bibinfo
  {author} {\bibfnamefont {H.~L.}\ \bibnamefont {Swinney}}, \bibinfo {author}
  {\bibfnamefont {A.}~\bibnamefont {Ardelea}}, \ and\ \bibinfo {author}
  {\bibfnamefont {G.~F.}\ \bibnamefont {Carey}},\ }\href {\doibase
  10.1103/PhysRevLett.84.4240} {\bibfield  {journal} {\bibinfo  {journal}
  {Physical Review Letters}\ }\textbf {\bibinfo {volume} {84}},\ \bibinfo
  {pages} {4240} (\bibinfo {year} {2000})}\BibitemShut {NoStop}%
\bibitem [{\citenamefont {Staliunas}\ \emph {et~al.}(2002)\citenamefont
  {Staliunas}, \citenamefont {Longhi},\ and\ \citenamefont
  {de~Valc{\'{a}}rcel}}]{Staliunas2002}%
  \BibitemOpen
  \bibfield  {author} {\bibinfo {author} {\bibfnamefont {K.}~\bibnamefont
  {Staliunas}}, \bibinfo {author} {\bibfnamefont {S.}~\bibnamefont {Longhi}}, \
  and\ \bibinfo {author} {\bibfnamefont {G.~J.}\ \bibnamefont
  {de~Valc{\'{a}}rcel}},\ }\href {\doibase 10.1103/PhysRevLett.89.210406}
  {\bibfield  {journal} {\bibinfo  {journal} {Physical Review Letters}\
  }\textbf {\bibinfo {volume} {89}},\ \bibinfo {pages} {210406} (\bibinfo
  {year} {2002})}\BibitemShut {NoStop}%
\bibitem [{\citenamefont {Engels}\ \emph {et~al.}(2007)\citenamefont {Engels},
  \citenamefont {Atherton},\ and\ \citenamefont {Hoefer}}]{Engels2007}%
  \BibitemOpen
  \bibfield  {author} {\bibinfo {author} {\bibfnamefont {P.}~\bibnamefont
  {Engels}}, \bibinfo {author} {\bibfnamefont {C.}~\bibnamefont {Atherton}}, \
  and\ \bibinfo {author} {\bibfnamefont {M.~A.}\ \bibnamefont {Hoefer}},\
  }\href {\doibase 10.1103/PhysRevLett.98.095301} {\bibfield  {journal}
  {\bibinfo  {journal} {Physical Review Letters}\ }\textbf {\bibinfo {volume}
  {98}},\ \bibinfo {pages} {095301} (\bibinfo {year} {2007})}\BibitemShut
  {NoStop}%
\bibitem [{\citenamefont {Szwaj}\ \emph {et~al.}(1998)\citenamefont {Szwaj},
  \citenamefont {Bielawski}, \citenamefont {Derozier},\ and\ \citenamefont
  {Erneux}}]{Szwaj1998}%
  \BibitemOpen
  \bibfield  {author} {\bibinfo {author} {\bibfnamefont {C.}~\bibnamefont
  {Szwaj}}, \bibinfo {author} {\bibfnamefont {S.}~\bibnamefont {Bielawski}},
  \bibinfo {author} {\bibfnamefont {D.}~\bibnamefont {Derozier}}, \ and\
  \bibinfo {author} {\bibfnamefont {T.}~\bibnamefont {Erneux}},\ }\href
  {\doibase 10.1103/PhysRevLett.80.3968} {\bibfield  {journal} {\bibinfo
  {journal} {Physical Review Letters}\ }\textbf {\bibinfo {volume} {80}},\
  \bibinfo {pages} {3968} (\bibinfo {year} {1998})}\BibitemShut {NoStop}%
\bibitem [{\citenamefont {Piccardo}\ and\ \citenamefont
  {Tubino}(2008)}]{Piccardo2008}%
  \BibitemOpen
  \bibfield  {author} {\bibinfo {author} {\bibfnamefont {G.}~\bibnamefont
  {Piccardo}}\ and\ \bibinfo {author} {\bibfnamefont {F.}~\bibnamefont
  {Tubino}},\ }\href {\doibase 10.1016/j.jsv.2007.09.008} {\bibfield  {journal}
  {\bibinfo  {journal} {Journal of Sound and Vibration}\ }\textbf {\bibinfo
  {volume} {311}},\ \bibinfo {pages} {353} (\bibinfo {year}
  {2008})}\BibitemShut {NoStop}%
\bibitem [{\citenamefont {Conforti}\ \emph {et~al.}(2014)\citenamefont
  {Conforti}, \citenamefont {Mussot}, \citenamefont {Kudlinski},\ and\
  \citenamefont {Trillo}}]{Conforti2014}%
  \BibitemOpen
  \bibfield  {author} {\bibinfo {author} {\bibfnamefont {M.}~\bibnamefont
  {Conforti}}, \bibinfo {author} {\bibfnamefont {A.}~\bibnamefont {Mussot}},
  \bibinfo {author} {\bibfnamefont {A.}~\bibnamefont {Kudlinski}}, \ and\
  \bibinfo {author} {\bibfnamefont {S.}~\bibnamefont {Trillo}},\ }\href
  {\doibase 10.1364/OL.39.004200} {\bibfield  {journal} {\bibinfo  {journal}
  {Optics Letters}\ }\textbf {\bibinfo {volume} {39}},\ \bibinfo {pages} {4200}
  (\bibinfo {year} {2014})}\BibitemShut {NoStop}%
\bibitem [{\citenamefont {Tarasov}\ \emph {et~al.}(2016)\citenamefont
  {Tarasov}, \citenamefont {Perego}, \citenamefont {Churkin}, \citenamefont
  {Staliunas},\ and\ \citenamefont {Turitsyn}}]{Tarasov2016}%
  \BibitemOpen
  \bibfield  {author} {\bibinfo {author} {\bibfnamefont {N.}~\bibnamefont
  {Tarasov}}, \bibinfo {author} {\bibfnamefont {A.~M.}\ \bibnamefont {Perego}},
  \bibinfo {author} {\bibfnamefont {D.~V.}\ \bibnamefont {Churkin}}, \bibinfo
  {author} {\bibfnamefont {K.}~\bibnamefont {Staliunas}}, \ and\ \bibinfo
  {author} {\bibfnamefont {S.~K.}\ \bibnamefont {Turitsyn}},\ }\href {\doibase
  10.1038/ncomms12441} {\bibfield  {journal} {\bibinfo  {journal} {Nature
  Communications}\ }\textbf {\bibinfo {volume} {7}},\ \bibinfo {pages} {12441}
  (\bibinfo {year} {2016})}\BibitemShut {NoStop}%
\bibitem [{\citenamefont {Perego}\ \emph {et~al.}(2018)\citenamefont {Perego},
  \citenamefont {Smirnov}, \citenamefont {Staliunas}, \citenamefont {Churkin},\
  and\ \citenamefont {Wabnitz}}]{Perego2018}%
  \BibitemOpen
  \bibfield  {author} {\bibinfo {author} {\bibfnamefont {A.}~\bibnamefont
  {Perego}}, \bibinfo {author} {\bibfnamefont {S.}~\bibnamefont {Smirnov}},
  \bibinfo {author} {\bibfnamefont {K.}~\bibnamefont {Staliunas}}, \bibinfo
  {author} {\bibfnamefont {D.}~\bibnamefont {Churkin}}, \ and\ \bibinfo
  {author} {\bibfnamefont {S.}~\bibnamefont {Wabnitz}},\ }\href {\doibase
  10.1103/PhysRevLett.120.213902} {\bibfield  {journal} {\bibinfo  {journal}
  {Physical Review Letters}\ }\textbf {\bibinfo {volume} {120}},\ \bibinfo
  {pages} {213902} (\bibinfo {year} {2018})}\BibitemShut {NoStop}%
\bibitem [{\citenamefont {Abdullaev}\ \emph {et~al.}(1997)\citenamefont
  {Abdullaev}, \citenamefont {Darmanyan}, \citenamefont {Bischoff},\ and\
  \citenamefont {S{\o}rensen}}]{Abdullaev1997}%
  \BibitemOpen
  \bibfield  {author} {\bibinfo {author} {\bibfnamefont {F.~K.}\ \bibnamefont
  {Abdullaev}}, \bibinfo {author} {\bibfnamefont {S.~A.}\ \bibnamefont
  {Darmanyan}}, \bibinfo {author} {\bibfnamefont {S.}~\bibnamefont {Bischoff}},
  \ and\ \bibinfo {author} {\bibfnamefont {M.~P.}\ \bibnamefont
  {S{\o}rensen}},\ }\href {\doibase 10.1364/JOSAB.14.000027} {\bibfield
  {journal} {\bibinfo  {journal} {Journal of the Optical Society of America B}\
  }\textbf {\bibinfo {volume} {14}},\ \bibinfo {pages} {27} (\bibinfo {year}
  {1997})}\BibitemShut {NoStop}%
\bibitem [{\citenamefont {Staliunas}\ \emph {et~al.}(2013)\citenamefont
  {Staliunas}, \citenamefont {Hang},\ and\ \citenamefont
  {Konotop}}]{Staliunas2013}%
  \BibitemOpen
  \bibfield  {author} {\bibinfo {author} {\bibfnamefont {K.}~\bibnamefont
  {Staliunas}}, \bibinfo {author} {\bibfnamefont {C.}~\bibnamefont {Hang}}, \
  and\ \bibinfo {author} {\bibfnamefont {V.~V.}\ \bibnamefont {Konotop}},\
  }\href {\doibase 10.1103/PhysRevA.88.023846} {\bibfield  {journal} {\bibinfo
  {journal} {Physical Review A}\ }\textbf {\bibinfo {volume} {88}},\ \bibinfo
  {pages} {023846} (\bibinfo {year} {2013})}\BibitemShut {NoStop}%
\bibitem [{\citenamefont {Armstrong}\ \emph {et~al.}(1962)\citenamefont
  {Armstrong}, \citenamefont {Bloembergen}, \citenamefont {Ducuing},\ and\
  \citenamefont {Pershan}}]{Armstrong1962}%
  \BibitemOpen
  \bibfield  {author} {\bibinfo {author} {\bibfnamefont {J.~A.}\ \bibnamefont
  {Armstrong}}, \bibinfo {author} {\bibfnamefont {N.}~\bibnamefont
  {Bloembergen}}, \bibinfo {author} {\bibfnamefont {J.}~\bibnamefont
  {Ducuing}}, \ and\ \bibinfo {author} {\bibfnamefont {P.~S.}\ \bibnamefont
  {Pershan}},\ }\href {\doibase 10.1103/PhysRev.127.1918} {\bibfield  {journal}
  {\bibinfo  {journal} {Physical Review}\ }\textbf {\bibinfo {volume} {127}},\
  \bibinfo {pages} {1918} (\bibinfo {year} {1962})}\BibitemShut {NoStop}%
\bibitem [{\citenamefont {Feng}\ \emph {et~al.}(1980)\citenamefont {Feng},
  \citenamefont {Ming}, \citenamefont {Hong}, \citenamefont {Yang},
  \citenamefont {Zhu}, \citenamefont {Yang},\ and\ \citenamefont
  {Wang}}]{Feng1980}%
  \BibitemOpen
  \bibfield  {author} {\bibinfo {author} {\bibfnamefont {D.}~\bibnamefont
  {Feng}}, \bibinfo {author} {\bibfnamefont {N.}~\bibnamefont {Ming}}, \bibinfo
  {author} {\bibfnamefont {J.}~\bibnamefont {Hong}}, \bibinfo {author}
  {\bibfnamefont {Y.}~\bibnamefont {Yang}}, \bibinfo {author} {\bibfnamefont
  {J.}~\bibnamefont {Zhu}}, \bibinfo {author} {\bibfnamefont {Z.}~\bibnamefont
  {Yang}}, \ and\ \bibinfo {author} {\bibfnamefont {Y.}~\bibnamefont {Wang}},\
  }\href {\doibase 10.1063/1.92035} {\bibfield  {journal} {\bibinfo  {journal}
  {Applied Physics Letters}\ }\textbf {\bibinfo {volume} {37}},\ \bibinfo
  {pages} {607} (\bibinfo {year} {1980})}\BibitemShut {NoStop}%
\bibitem [{\citenamefont {Jundt}\ \emph {et~al.}(1991)\citenamefont {Jundt},
  \citenamefont {Magel}, \citenamefont {Fejer},\ and\ \citenamefont
  {Byer}}]{Jundt1991}%
  \BibitemOpen
  \bibfield  {author} {\bibinfo {author} {\bibfnamefont {D.~H.}\ \bibnamefont
  {Jundt}}, \bibinfo {author} {\bibfnamefont {G.~A.}\ \bibnamefont {Magel}},
  \bibinfo {author} {\bibfnamefont {M.~M.}\ \bibnamefont {Fejer}}, \ and\
  \bibinfo {author} {\bibfnamefont {R.~L.}\ \bibnamefont {Byer}},\ }\href
  {\doibase 10.1063/1.105929} {\bibfield  {journal} {\bibinfo  {journal}
  {Applied Physics Letters}\ }\textbf {\bibinfo {volume} {59}},\ \bibinfo
  {pages} {2657} (\bibinfo {year} {1991})}\BibitemShut {NoStop}%
\bibitem [{\citenamefont {Mizuuchi}\ and\ \citenamefont
  {Yamamoto}(1992)}]{Mizuuchi1992}%
  \BibitemOpen
  \bibfield  {author} {\bibinfo {author} {\bibfnamefont {K.}~\bibnamefont
  {Mizuuchi}}\ and\ \bibinfo {author} {\bibfnamefont {K.}~\bibnamefont
  {Yamamoto}},\ }\href {\doibase 10.1063/1.107317} {\bibfield  {journal}
  {\bibinfo  {journal} {Applied Physics Letters}\ }\textbf {\bibinfo {volume}
  {60}},\ \bibinfo {pages} {1283} (\bibinfo {year} {1992})}\BibitemShut
  {NoStop}%
\bibitem [{\citenamefont {Fejer}\ \emph {et~al.}(1992)\citenamefont {Fejer},
  \citenamefont {Magel}, \citenamefont {Jundt},\ and\ \citenamefont
  {Byer}}]{Bromberg1992}%
  \BibitemOpen
  \bibfield  {author} {\bibinfo {author} {\bibfnamefont {M.~M.}\ \bibnamefont
  {Fejer}}, \bibinfo {author} {\bibfnamefont {G.~A.}\ \bibnamefont {Magel}},
  \bibinfo {author} {\bibfnamefont {D.~H.}\ \bibnamefont {Jundt}}, \ and\
  \bibinfo {author} {\bibfnamefont {R.~L.}\ \bibnamefont {Byer}},\ }\href
  {\doibase 10.1109/3.161322} {\bibfield  {journal} {\bibinfo  {journal} {Ieee
  Journal of Quantum Electronics}\ }\textbf {\bibinfo {volume} {102}},\
  \bibinfo {pages} {2631} (\bibinfo {year} {1992})}\BibitemShut {NoStop}%
\bibitem [{\citenamefont {Bortz}\ \emph {et~al.}(1995)\citenamefont {Bortz},
  \citenamefont {Arbore},\ and\ \citenamefont {Fejer}}]{Bortz1995}%
  \BibitemOpen
  \bibfield  {author} {\bibinfo {author} {\bibfnamefont {M.~L.}\ \bibnamefont
  {Bortz}}, \bibinfo {author} {\bibfnamefont {M.~a.}\ \bibnamefont {Arbore}}, \
  and\ \bibinfo {author} {\bibfnamefont {M.~M.}\ \bibnamefont {Fejer}},\ }\href
  {\doibase 10.1364/OL.20.000049} {\bibfield  {journal} {\bibinfo  {journal}
  {Optics letters}\ }\textbf {\bibinfo {volume} {20}},\ \bibinfo {pages} {49}
  (\bibinfo {year} {1995})}\BibitemShut {NoStop}%
\bibitem [{\citenamefont {Myers}\ \emph {et~al.}(1995)\citenamefont {Myers},
  \citenamefont {Eckardt}, \citenamefont {Fejer}, \citenamefont {Byer},
  \citenamefont {Bosenberg},\ and\ \citenamefont {Pierce}}]{Myers1995}%
  \BibitemOpen
  \bibfield  {author} {\bibinfo {author} {\bibfnamefont {L.~E.}\ \bibnamefont
  {Myers}}, \bibinfo {author} {\bibfnamefont {R.~C.}\ \bibnamefont {Eckardt}},
  \bibinfo {author} {\bibfnamefont {M.~M.}\ \bibnamefont {Fejer}}, \bibinfo
  {author} {\bibfnamefont {R.~L.}\ \bibnamefont {Byer}}, \bibinfo {author}
  {\bibfnamefont {W.~R.}\ \bibnamefont {Bosenberg}}, \ and\ \bibinfo {author}
  {\bibfnamefont {J.~W.}\ \bibnamefont {Pierce}},\ }\href {\doibase
  10.1364/JOSAB.12.002102} {\bibfield  {journal} {\bibinfo  {journal} {Journal
  of the Optical Society of America B}\ }\textbf {\bibinfo {volume} {12}},\
  \bibinfo {pages} {2102} (\bibinfo {year} {1995})}\BibitemShut {NoStop}%
\bibitem [{\citenamefont {Tanzilli}\ \emph {et~al.}(2001)\citenamefont
  {Tanzilli}, \citenamefont {De~Riedmatten}, \citenamefont {Tittel},
  \citenamefont {Zbinden}, \citenamefont {Baldi}, \citenamefont {De~Micheli},
  \citenamefont {Ostrowsky},\ and\ \citenamefont {Gisin}}]{Tanzilli2001b}%
  \BibitemOpen
  \bibfield  {author} {\bibinfo {author} {\bibfnamefont {S.}~\bibnamefont
  {Tanzilli}}, \bibinfo {author} {\bibfnamefont {H.}~\bibnamefont
  {De~Riedmatten}}, \bibinfo {author} {\bibfnamefont {W.}~\bibnamefont
  {Tittel}}, \bibinfo {author} {\bibfnamefont {H.}~\bibnamefont {Zbinden}},
  \bibinfo {author} {\bibfnamefont {P.}~\bibnamefont {Baldi}}, \bibinfo
  {author} {\bibfnamefont {M.}~\bibnamefont {De~Micheli}}, \bibinfo {author}
  {\bibfnamefont {D.}~\bibnamefont {Ostrowsky}}, \ and\ \bibinfo {author}
  {\bibfnamefont {N.}~\bibnamefont {Gisin}},\ }\href {\doibase
  10.1049/el:20010009} {\bibfield  {journal} {\bibinfo  {journal} {Electronics
  Letters}\ }\textbf {\bibinfo {volume} {37}},\ \bibinfo {pages} {26} (\bibinfo
  {year} {2001})}\BibitemShut {NoStop}%
\bibitem [{\citenamefont {Banaszek}\ \emph {et~al.}(2001)\citenamefont
  {Banaszek}, \citenamefont {U'ren},\ and\ \citenamefont
  {Walmsley}}]{Banaszek2001}%
  \BibitemOpen
  \bibfield  {author} {\bibinfo {author} {\bibfnamefont {K.}~\bibnamefont
  {Banaszek}}, \bibinfo {author} {\bibfnamefont {A.~B.}\ \bibnamefont {U'ren}},
  \ and\ \bibinfo {author} {\bibfnamefont {I.~A.}\ \bibnamefont {Walmsley}},\
  }\href {\doibase 10.1364/OL.26.001367} {\bibfield  {journal} {\bibinfo
  {journal} {Optics letters}\ }\textbf {\bibinfo {volume} {26}},\ \bibinfo
  {pages} {1367} (\bibinfo {year} {2001})}\BibitemShut {NoStop}%
\bibitem [{\citenamefont {Xu}\ and\ \citenamefont {Zhu}(2012)}]{Xu2012d}%
  \BibitemOpen
  \bibfield  {author} {\bibinfo {author} {\bibfnamefont {P.}~\bibnamefont
  {Xu}}\ and\ \bibinfo {author} {\bibfnamefont {S.~N.}\ \bibnamefont {Zhu}},\
  }\href {\doibase 10.1063/1.4773457} {\bibfield  {journal} {\bibinfo
  {journal} {AIP Advances}\ }\textbf {\bibinfo {volume} {2}},\ \bibinfo {pages}
  {041401} (\bibinfo {year} {2012})}\BibitemShut {NoStop}%
\bibitem [{\citenamefont {Burnham}\ and\ \citenamefont
  {Weinberg}(1970)}]{Burnham1970b}%
  \BibitemOpen
  \bibfield  {author} {\bibinfo {author} {\bibfnamefont {D.~C.}\ \bibnamefont
  {Burnham}}\ and\ \bibinfo {author} {\bibfnamefont {D.~L.}\ \bibnamefont
  {Weinberg}},\ }\href {\doibase 10.1103/PhysRevLett.25.84} {\bibfield
  {journal} {\bibinfo  {journal} {Physical Review Letters}\ }\textbf {\bibinfo
  {volume} {25}},\ \bibinfo {pages} {84} (\bibinfo {year} {1970})}\BibitemShut
  {NoStop}%
\bibitem [{\citenamefont {Eisaman}\ \emph {et~al.}(2011)\citenamefont
  {Eisaman}, \citenamefont {Fan}, \citenamefont {Migdall},\ and\ \citenamefont
  {Polyakov}}]{Eisaman2011}%
  \BibitemOpen
  \bibfield  {author} {\bibinfo {author} {\bibfnamefont {M.~D.}\ \bibnamefont
  {Eisaman}}, \bibinfo {author} {\bibfnamefont {J.}~\bibnamefont {Fan}},
  \bibinfo {author} {\bibfnamefont {A.}~\bibnamefont {Migdall}}, \ and\
  \bibinfo {author} {\bibfnamefont {S.~V.}\ \bibnamefont {Polyakov}},\ }\href
  {\doibase 10.1063/1.3610677} {\bibfield  {journal} {\bibinfo  {journal}
  {Review of Scientific Instruments}\ }\textbf {\bibinfo {volume} {82}},\
  \bibinfo {pages} {071101} (\bibinfo {year} {2011})}\BibitemShut {NoStop}%
\bibitem [{\citenamefont {Couteau}(2018)}]{Couteau2018}%
  \BibitemOpen
  \bibfield  {author} {\bibinfo {author} {\bibfnamefont {C.}~\bibnamefont
  {Couteau}},\ }\href {\doibase 10.1080/00107514.2018.1488463} {\bibfield
  {journal} {\bibinfo  {journal} {Contemporary Physics}\ }\textbf {\bibinfo
  {volume} {59}},\ \bibinfo {pages} {291} (\bibinfo {year} {2018})}\BibitemShut
  {NoStop}%
\bibitem [{\citenamefont {Main}\ \emph {et~al.}(2016)\citenamefont {Main},
  \citenamefont {Mosley}, \citenamefont {Ding}, \citenamefont {Zhang},\ and\
  \citenamefont {Gorbach}}]{Main2016}%
  \BibitemOpen
  \bibfield  {author} {\bibinfo {author} {\bibfnamefont {P.}~\bibnamefont
  {Main}}, \bibinfo {author} {\bibfnamefont {P.~J.}\ \bibnamefont {Mosley}},
  \bibinfo {author} {\bibfnamefont {W.}~\bibnamefont {Ding}}, \bibinfo {author}
  {\bibfnamefont {L.}~\bibnamefont {Zhang}}, \ and\ \bibinfo {author}
  {\bibfnamefont {A.~V.}\ \bibnamefont {Gorbach}},\ }\href {\doibase
  10.1103/PhysRevA.94.063844} {\bibfield  {journal} {\bibinfo  {journal}
  {Physical Review A}\ }\textbf {\bibinfo {volume} {94}},\ \bibinfo {pages}
  {063844} (\bibinfo {year} {2016})}\BibitemShut {NoStop}%
\bibitem [{\citenamefont {Mandel}\ and\ \citenamefont
  {Wolf}(1995)}]{Mandel1995}%
  \BibitemOpen
  \bibfield  {author} {\bibinfo {author} {\bibfnamefont {L.}~\bibnamefont
  {Mandel}}\ and\ \bibinfo {author} {\bibfnamefont {E.}~\bibnamefont {Wolf}},\
  }\href {\doibase 10.1119/1.18450} {\emph {\bibinfo {title} {{Optical
  Coherence and Quantum Optics}}}}\ (\bibinfo  {publisher} {Cambridge
  university press},\ \bibinfo {year} {1995})\BibitemShut {NoStop}%
\bibitem [{\citenamefont {Rubin}(1996)}]{Rubin1996}%
  \BibitemOpen
  \bibfield  {author} {\bibinfo {author} {\bibfnamefont {M.~H.}\ \bibnamefont
  {Rubin}},\ }\href {\doibase 10.1103/PhysRevA.54.5349} {\bibfield  {journal}
  {\bibinfo  {journal} {Physical Review A}\ }\textbf {\bibinfo {volume} {54}},\
  \bibinfo {pages} {5349} (\bibinfo {year} {1996})}\BibitemShut {NoStop}%
\bibitem [{\citenamefont {Grice}\ and\ \citenamefont
  {Walmsley}(1997)}]{Grice1997a}%
  \BibitemOpen
  \bibfield  {author} {\bibinfo {author} {\bibfnamefont {W.~P.}\ \bibnamefont
  {Grice}}\ and\ \bibinfo {author} {\bibfnamefont {I.~A.}\ \bibnamefont
  {Walmsley}},\ }\href {\doibase 10.1103/PhysRevA.56.1627} {\bibfield
  {journal} {\bibinfo  {journal} {Phys. Rev. A}\ }\textbf {\bibinfo {volume}
  {56}},\ \bibinfo {pages} {1627} (\bibinfo {year} {1997})}\BibitemShut
  {NoStop}%
\bibitem [{\citenamefont {Yang}\ \emph {et~al.}(2008)\citenamefont {Yang},
  \citenamefont {Liscidini},\ and\ \citenamefont {Sipe}}]{Yang2008a}%
  \BibitemOpen
  \bibfield  {author} {\bibinfo {author} {\bibfnamefont {Z.}~\bibnamefont
  {Yang}}, \bibinfo {author} {\bibfnamefont {M.}~\bibnamefont {Liscidini}}, \
  and\ \bibinfo {author} {\bibfnamefont {J.~E.}\ \bibnamefont {Sipe}},\ }\href
  {\doibase 10.1103/PhysRevA.77.033808} {\bibfield  {journal} {\bibinfo
  {journal} {Physical Review A}\ }\textbf {\bibinfo {volume} {77}},\ \bibinfo
  {pages} {033808} (\bibinfo {year} {2008})}\BibitemShut {NoStop}%
\bibitem [{\citenamefont {Agrawal}(2013)}]{agrawal}%
  \BibitemOpen
  \bibfield  {author} {\bibinfo {author} {\bibfnamefont {G.~P.}\ \bibnamefont
  {Agrawal}},\ }\href@noop {} {\emph {\bibinfo {title} {{Nonlinear Fiber
  Optics}}}},\ \bibinfo {edition} {5th}\ ed.\ (\bibinfo  {publisher} {Academic
  Press},\ \bibinfo {year} {2013})\BibitemShut {NoStop}%
\bibitem [{\citenamefont {Arnol'd}(1983)}]{Arnold1983}%
  \BibitemOpen
  \bibfield  {author} {\bibinfo {author} {\bibfnamefont {V.~I.}\ \bibnamefont
  {Arnol'd}},\ }\href {\doibase 10.1070/RM1983v038n04ABEH004210} {\bibfield
  {journal} {\bibinfo  {journal} {Russian Mathematical Surveys}\ }\textbf
  {\bibinfo {volume} {38}},\ \bibinfo {pages} {215} (\bibinfo {year}
  {1983})}\BibitemShut {NoStop}%
\bibitem [{\citenamefont {Ecke}\ \emph {et~al.}(1989)\citenamefont {Ecke},
  \citenamefont {Farmer},\ and\ \citenamefont {Umberger}}]{Ecke1989}%
  \BibitemOpen
  \bibfield  {author} {\bibinfo {author} {\bibfnamefont {R.~E.}\ \bibnamefont
  {Ecke}}, \bibinfo {author} {\bibfnamefont {J.~D.}\ \bibnamefont {Farmer}}, \
  and\ \bibinfo {author} {\bibfnamefont {D.~K.}\ \bibnamefont {Umberger}},\
  }\href {\doibase 10.1088/0951-7715/2/2/001} {\bibfield  {journal} {\bibinfo
  {journal} {Nonlinearity}\ }\textbf {\bibinfo {volume} {2}},\ \bibinfo {pages}
  {175} (\bibinfo {year} {1989})}\BibitemShut {NoStop}%
\bibitem [{\citenamefont {Hindmarsh}(1982)}]{Hindmarsh1982}%
  \BibitemOpen
  \bibfield  {author} {\bibinfo {author} {\bibfnamefont {A.}~\bibnamefont
  {Hindmarsh}},\ }\href {\doibase 10.1109/MCS.1982.1103756} {\bibfield
  {journal} {\bibinfo  {journal} {IEEE Control Systems Magazine}\ }\textbf
  {\bibinfo {volume} {2}},\ \bibinfo {pages} {24} (\bibinfo {year}
  {1982})}\BibitemShut {NoStop}%
\bibitem [{\citenamefont {Perez-Leija}\ \emph {et~al.}(2012)\citenamefont
  {Perez-Leija}, \citenamefont {Keil}, \citenamefont {Szameit}, \citenamefont
  {Abouraddy}, \citenamefont {Moya-Cessa},\ and\ \citenamefont
  {Christodoulides}}]{Perez-Leija2012}%
  \BibitemOpen
  \bibfield  {author} {\bibinfo {author} {\bibfnamefont {A.}~\bibnamefont
  {Perez-Leija}}, \bibinfo {author} {\bibfnamefont {R.}~\bibnamefont {Keil}},
  \bibinfo {author} {\bibfnamefont {A.}~\bibnamefont {Szameit}}, \bibinfo
  {author} {\bibfnamefont {A.~F.}\ \bibnamefont {Abouraddy}}, \bibinfo {author}
  {\bibfnamefont {H.}~\bibnamefont {Moya-Cessa}}, \ and\ \bibinfo {author}
  {\bibfnamefont {D.~N.}\ \bibnamefont {Christodoulides}},\ }\href {\doibase
  10.1103/PhysRevA.85.013848} {\bibfield  {journal} {\bibinfo  {journal}
  {Physical Review A}\ }\textbf {\bibinfo {volume} {85}},\ \bibinfo {pages}
  {013848} (\bibinfo {year} {2012})}\BibitemShut {NoStop}%
\bibitem [{\citenamefont {Sukhorukov}\ \emph {et~al.}(2013)\citenamefont
  {Sukhorukov}, \citenamefont {Solntsev},\ and\ \citenamefont
  {Sipe}}]{Sukhorukov2013a}%
  \BibitemOpen
  \bibfield  {author} {\bibinfo {author} {\bibfnamefont {A.~A.}\ \bibnamefont
  {Sukhorukov}}, \bibinfo {author} {\bibfnamefont {A.~S.}\ \bibnamefont
  {Solntsev}}, \ and\ \bibinfo {author} {\bibfnamefont {J.~E.}\ \bibnamefont
  {Sipe}},\ }\href {\doibase 10.1103/PhysRevA.87.053823} {\bibfield  {journal}
  {\bibinfo  {journal} {Physical Review A}\ }\textbf {\bibinfo {volume} {87}},\
  \bibinfo {pages} {053823} (\bibinfo {year} {2013})}\BibitemShut {NoStop}%
\bibitem [{\citenamefont {Izrailev}\ \emph {et~al.}(2012)\citenamefont
  {Izrailev}, \citenamefont {Krokhin},\ and\ \citenamefont
  {Makarov}}]{Izrailev2012}%
  \BibitemOpen
  \bibfield  {author} {\bibinfo {author} {\bibfnamefont {F.}~\bibnamefont
  {Izrailev}}, \bibinfo {author} {\bibfnamefont {A.}~\bibnamefont {Krokhin}}, \
  and\ \bibinfo {author} {\bibfnamefont {N.}~\bibnamefont {Makarov}},\ }\href
  {\doibase 10.1016/J.PHYSREP.2011.11.002} {\bibfield  {journal} {\bibinfo
  {journal} {Physics Reports}\ }\textbf {\bibinfo {volume} {512}},\ \bibinfo
  {pages} {125} (\bibinfo {year} {2012})}\BibitemShut {NoStop}%
\bibitem [{\citenamefont {Kumar}(1990)}]{Kumar1990}%
  \BibitemOpen
  \bibfield  {author} {\bibinfo {author} {\bibfnamefont {P.}~\bibnamefont
  {Kumar}},\ }\href {\doibase 10.1364/OL.15.001476} {\bibfield  {journal}
  {\bibinfo  {journal} {Optics Letters}\ }\textbf {\bibinfo {volume} {15}},\
  \bibinfo {pages} {1476} (\bibinfo {year} {1990})}\BibitemShut {NoStop}%
\bibitem [{\citenamefont {Wright}\ \emph {et~al.}(2018)\citenamefont {Wright},
  \citenamefont {Francis-Jones}, \citenamefont {Gawith}, \citenamefont
  {Becker}, \citenamefont {Ledingham}, \citenamefont {Smith}, \citenamefont
  {Nunn}, \citenamefont {Mosley}, \citenamefont {Brecht},\ and\ \citenamefont
  {Walmsley}}]{Wright2018}%
  \BibitemOpen
  \bibfield  {author} {\bibinfo {author} {\bibfnamefont {T.~A.}\ \bibnamefont
  {Wright}}, \bibinfo {author} {\bibfnamefont {R.~J.~A.}\ \bibnamefont
  {Francis-Jones}}, \bibinfo {author} {\bibfnamefont {C.~B.~E.}\ \bibnamefont
  {Gawith}}, \bibinfo {author} {\bibfnamefont {J.~N.}\ \bibnamefont {Becker}},
  \bibinfo {author} {\bibfnamefont {P.~M.}\ \bibnamefont {Ledingham}}, \bibinfo
  {author} {\bibfnamefont {P.~G.~R.}\ \bibnamefont {Smith}}, \bibinfo {author}
  {\bibfnamefont {J.}~\bibnamefont {Nunn}}, \bibinfo {author} {\bibfnamefont
  {P.~J.}\ \bibnamefont {Mosley}}, \bibinfo {author} {\bibfnamefont
  {B.}~\bibnamefont {Brecht}}, \ and\ \bibinfo {author} {\bibfnamefont {I.~A.}\
  \bibnamefont {Walmsley}},\ }\href@noop {} {\bibfield  {journal} {\bibinfo
  {journal} {Phys. Rev. Applied}\ }\textbf {\bibinfo {volume} {10}},\ \bibinfo
  {pages} {044012} (\bibinfo {year} {2018})}\BibitemShut {NoStop}%
\end{thebibliography}%

\end{document}